# Unified statistical distribution of quantum particles and Supersymmetry

Ahmad Adel Abutaleb

Department of Mathematics, Faculty of science, University of Mansoura, Elmansoura 35516, Egypt

**Abstract**

In this paper we propose a unified statistics of Bose-Einstein and Fermi-Dirac statistics by suggesting that every particle can be associated with matter or fundamental forces with certain probability. The main Justification for this proposal is the possibility of extension of the spin-statistics theory to include a hypothetical quantum particles have fractional spin. The concept of Supersymmetry can be related to this unified statistics.

## Introduction

The well-known distributions which describe how the number of indistinguishable fermions or bosons in different energy states varied with the energy of these states are Fermi-Dirac and Bose-Einstein distributions respectively [1]. Although all experimentally observed particles are either bosonic or Fermionic, the general principles of quantum mechanics do not prevent the existence of some objects obeying intermediate statistics [8]. In fact there is a huge amount of work of Interpolations between Bose-Einstein and Fermi-Dirac statistics, mostly triggered by the quantum hall effect and anyonic statistics [2,3,4,5]. The first work on this topic was made by Gentile [6], where his idea is based on Generalizing the Pauli exclusion principle i.e. based on the assumption that the maximum occupation number can have undetermined integer $0 < n < \infty$. Another approach is to introduce a parameter ($\alpha$) valued from 0 to 1 in the expression of the number of quantum states [7]. Also, Fractional statistics can be achieved by analyzing the symmetry properties of the wave function. The wave function change a phase factor when identical particles exchange, and the phase factor can be -1 or +1 related to Fermions and Bosons. The concepts of anyone and Fractional statistics appears when this result is generalized to an arbitrary phase factor $e^{i\theta}$ and this work was done in [3]. In this paper we introduce another intermediate statistics between Fermions and Bosons. Our approach is based on suggesting that every particle has a mixture properties of both Fermions and Bosons. This is equivalent to say that every particle is a bosonic with certain probability $(1-P)$ or Fermionic with probability $(P)$.

**Unified Statistics**

Suppose that we have a number of energy levels, labeled by index $(i)$, each level has energy $(\epsilon_i)$ and contain a total $(n_i)$ of particles and suppose that each level contain $(g_i)$ distnistic sublevels, all of which have the same energy and which are distinguishable. The well known statistical distributions if those particles were fermions or bosons are respectively [8], Fermi-Dirac distribution

$$n_i = \frac{g_i}{e^{(\epsilon_i-\mu)/KT}+1} \quad (1)$$

and Bose-Einstein distribution

$$n_i = \frac{g_i}{e^{(\epsilon_i-\mu)/KT}-1} \quad (2).$$

According to the standard model, all elementary particles are either bosons or fermions and the spin statistics theorem identifies the resulting quantum statistics that differentiates between them. According to this methodology, particles normally associated with matter are fermions, they have half-integer spin and are divided into twelve flavors. On the other hand the particles associated with fundamental forces are bosons and they have integer spin [10]. As wave-particle duality postulates that all particles exhibit both wave and particles properties, we assume that every quantum particle has properties of both fermions and bosons and may associate with matter or fundamental forces with certain probability. Therefore we want to find the form of statistical distribution if any of quantum particles can be a fermion with probability $(p)$ or a boson with probability $(1-p)$. This is equivalent to say that if the total number of particles is $(n_i)$ then we have $(pn_i)$ fermions and $(1-p)n_i$ bosons. We can write the number of total ways of distributing $(pn_i)$ fermions and $(1-p)n_i$ bosons among $(g_i)$ sublevels as

$$w_a = \prod_i \frac{g_i!}{pn_i! \, g_i - pn_i!} * \frac{(g_i+(1-p)n_i-1)!}{((1-p)n_i! \, g_i-1!)} . \tag{3}$$

As the usual way of deriving of both Femi and Bose statistics we want to know what is the value of $(n_i)$ which maximize $(w_a)$ under the condition of preserving the total energy and the total number of particles. We form the function

$$f_a(n_i) = \ln(w_a) + \alpha(N - \sum_i n_i) + \beta(E - \sum_i n_i \epsilon_i) \tag{4}$$

where $(\alpha, \beta)$ are Lagrange multipliers ,$(N)$ is the total number of particles and $(E)$ is the total energy. Using Stirling approximation for the factorials, taking the derivative with respect to $n_i$ , setting the result to zero, and solving for $n_i$ yields

$$\left(\frac{g_i}{(1-p)n_i} + 1\right)^{1-p} \left(\frac{g_i}{pn_i} - 1\right)^p = e^{\alpha + \beta \epsilon_i} . \tag{5}$$

It can be shown thermodynamically that [9]

$$\alpha = \frac{-\mu}{KT} \quad \text{and} \quad \beta = \frac{1}{KT} \tag{6}$$

where $(T)$ is the temperature ,$(K)$ is Boltzmann constant and $(\mu)$ is the chemical potential , so finally we have the unified statistics

$$\left(\frac{g_i}{(1-p)n_i} + 1\right)^{1-p} \left(\frac{g_i}{pn_i} - 1\right)^p = e^{(\epsilon_i - \mu)/KT} . \tag{7}$$

### Supersymmetry

Now , the new parameter $(p)$ is the probability that the particle exhibits the behavior of fermion. if we take the limit of distribution (7) when $p \to 1^-$ (limit from the lift) then we recover Fermi-Dirac distribution as follow

$\lim_{p \to 1^-} (\frac{g_i}{(1-p)n_i} + 1)^{1-p} (\frac{g_i}{pn_i} - 1)^p = (\frac{g_i}{n_i} - 1) = e^{(\epsilon_i - \mu)/KT}$ and by the same way we can recover Bose-Einstein distribution by taking the limit of distribution (7) when $p \to 0^+$. In distribution (7), all quantum particles divided not only to bosons and fermions but also divided to infinitely subclasses of particles. In Supersymmetry, each fermion is associated with some boson called (*superpartner*), so we can associate each subclass ( determined by the parameter $p$ ) by another (determined by the parameter $(1 - p)$). We will discuss this issue.

Let $\frac{g_i}{n_i} = x$ , $e^{(\epsilon_i - \mu)/KT} = r$ and suppose also that $r > 1$, so distribution (7) simplify to

$$(\frac{x}{1-p} + 1)^{1-p} (\frac{x}{p} - 1)^p = r . \qquad (8)$$

First, we want to stress that although equation (8) led to negative and complex roots according to the value of the parameter $p$, Equation (8) have one and only one positive real root (for $r > 1$) . Consider some quantum particle obey the unified statistics (8) with certain probability $(p)$ then we can define the *superpartner* of this quantum particle as the particle which obey (8) with the transformation of the parameter $p \to 1 - p$ i.e. obey the following statistics

$$(\frac{x}{1-p} - 1)^{1-p} (\frac{x}{p} + 1)^p = r . \qquad (9)$$

In other word, if the particle $A$ obey the unified distribution (8) with $p = \frac{1}{4}$ then the *superpartner* of $A$ obey the unified distribution (8) with $p = \frac{3}{4}$ (or equivalently obey the distribution (9) with $p = \frac{1}{4}$) . As a numerical examples let $r = 2$ , so for pure fermion we have $x = r + 1 = 3$ and for pure boson we have $x = r - 1 = 1$. We plot the real roots obtained from the statistics of particle and its *superpartner* for some special values of the parameter $p$ .

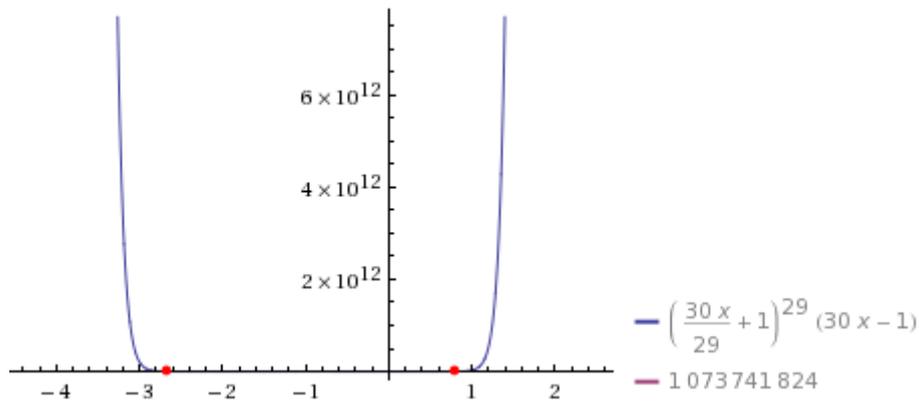

Fig (1) (statistics of some quantum particle $A$) (The real roots are x=-2.66831, x=0.809738) , $p = \frac{1}{30}$ (close to boson) .

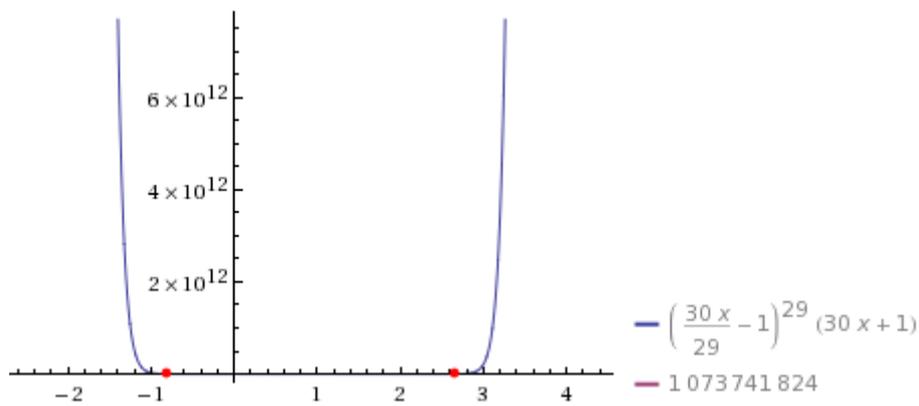

Fig (2) (statistics of the *superpartner* of particle $A$) (The real roots are x=2,66831, x=-0.809738 ) ,$p = \frac{29}{30}$ (close to fermion) .

In Fig (1), the only positive real root when $p = \frac{1}{30}$ is $x = 0.809738$ ($x = 1$ for pure boson). Notice that $x \to 1$ as long as $p \to 0$ (and the other negative rote $x \to -3$ as long as $p \to 0$. However, in some cases, the distribution (8) or (9) have no negative real roots .

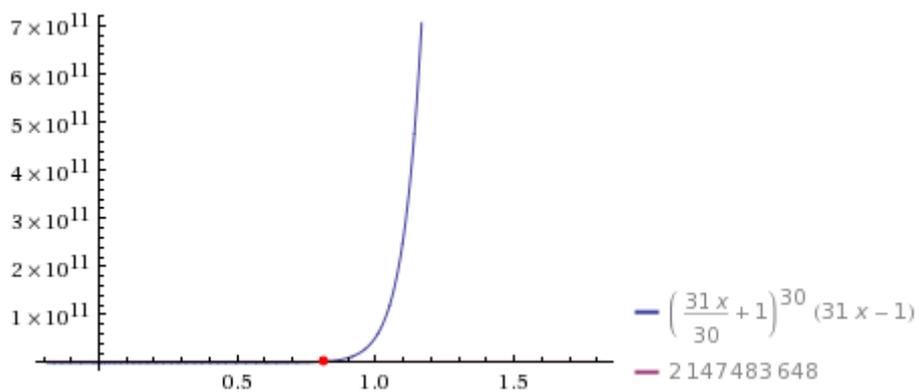

Fig (3) (the only real root $x = 0.813358$) , $p = \frac{1}{31}$.

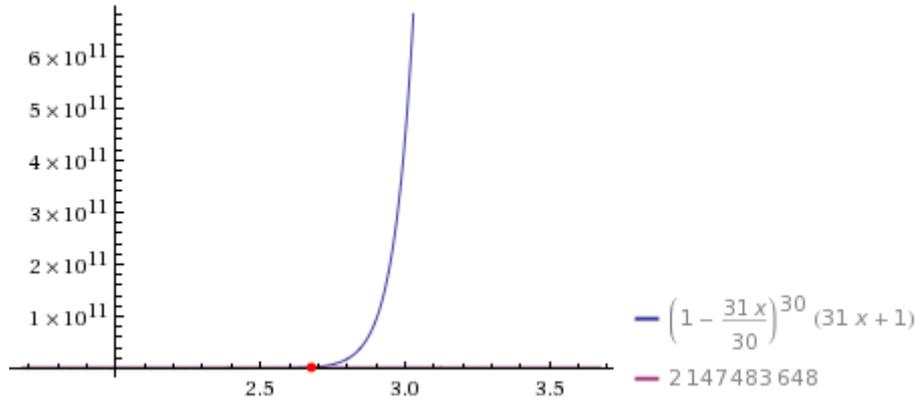

Fig (4) (the only real root $x = 2.67652$) , $p = \frac{30}{31}$.

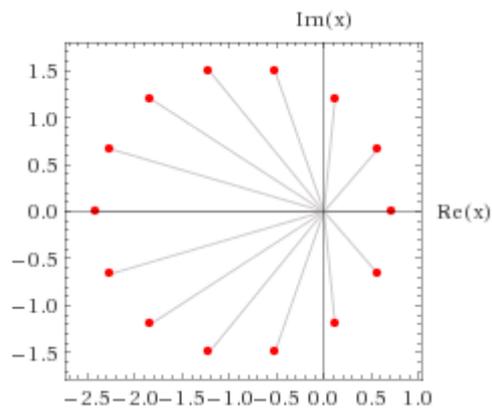

Fig (5) (the roots in complex plane)(statistics of some particle $A$) , $p = \frac{1}{14}$.

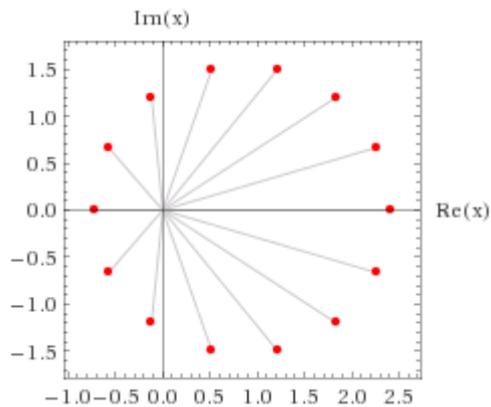

Fig (6) (the roots in complex plane)(statistics of *superpartner* of $A$), $p = \frac{13}{14}$.

Another feature of the symmetries between a quantum particle and its *superpartner* can be obtained by finding the transformations which transfer the distribution (8) of some quantum particle to the distribution (9) of its *superpartner*. In other word, by rewriting distribution (9) as

$$(\frac{y}{1-p} - 1)^{1-p} (\frac{y}{p} + 1)^p = r \qquad (10)$$

we want to determine $y = y(x)$ which transform the distribution (10) to its *superpartner* distribution (8). We plot the following figures for some values of $p$.

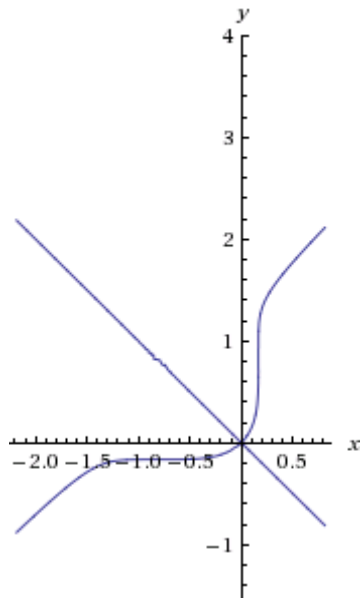

Fig (7) (relation between $x$ (statistics of some quantum particles) and $y$ (statistics of the *superpartner* of the quantum particle), $= \frac{1}{6}$.

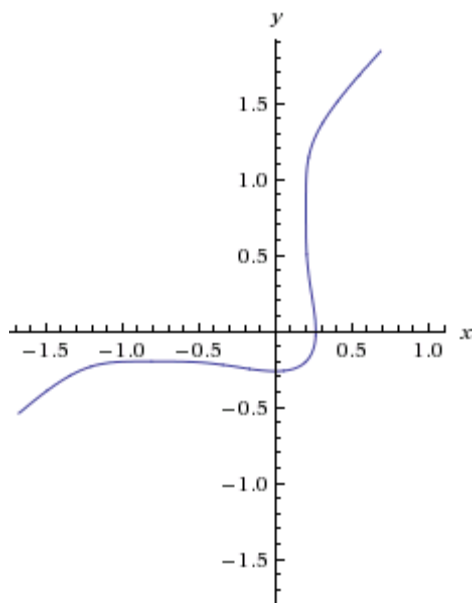

Fig (8) (relation between $x$ and $y$), $p = \frac{1}{5}$.

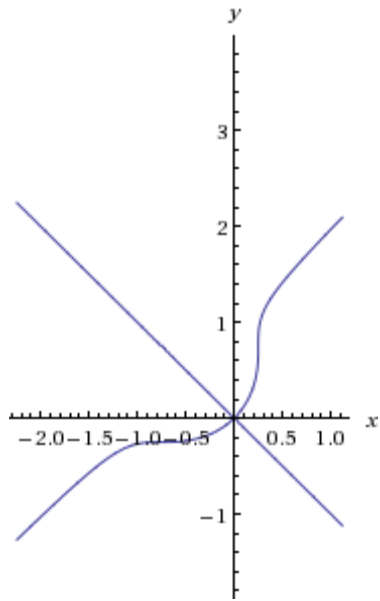

Fig (9) (relation between $x$ and $y$), $p = \frac{1}{4}$ .

**Conclusion**

The unified statistical distribution of all quantum partcles has been derived. Some of the symmetries feature of this distribution has been shown. We hope that we will find some applications of this distribution in the fractional quantum hall effect.

## Reference

[1] L. P. Kadanoff, *Statistical Physics*, World Scientific (2000).

[2] G. A. Goldin, R. Menikoff, D. H. Sharp, *J.Math Phys,* vol **21**, pp. 650-664 (1980).

[3] F. Wiliczek, *Phys. Rev. lett*, vol **49**, no 14, pp. 957-959 (1982)..

[4] M. V. Medvedev. *Phys. Rev. Lett*, vol **78**, no 22, pp. 4174-4150 (1997).

[5] F. D. M. Haldane, *Phys. Rev. Lett*, vol **67**, no 8, pp. 937-940 (1991).

[6] G. Gentile, *Nuovo Cimento*, vol **17**, pp. 493 (1940).

[7] Y. S. Wu , Phys. Rev. Lett, vol 73, no 7, pp. 922-925 (1994).

[8]  A. M. L. Messiah, O. W. Greenberg, *Phys. Rev*, vol **136**, p.p 248-267 (1964)

[9]A. H. Carter, *Classical and statistical Thermodynamics*, upper saddle River, N. J. Printice Hall ( 2001).

 [10] M. J. G. Veltmann, *Facts and mysteries in elementary particle physics*, world scientific (2003).